# Using YOLO v7 to Detect Kidney in Magnetic Resonance Imaging

Pouria Yazdian Anari[1], Fiona Obiezu [1], Nathan Lay[2], Fatemeh Dehghani Firouzabadi[1], Aditi Chaurasia[3], Mahshid Golagha [1], Shiva Singh[1], Fatemeh Homayounieh[1], Aryan Zahergivar [1], Stephanie Harmon[2], Evrim Turkbey[1], Rabindra Gautam[3], Kevin Ma[2], Maria Merino[4], Elizabeth C. Jones[1], Mark W. Ball[3], W. Marston Linehan[3], Baris Turkbey[2], Ashkan A. Malayeri[1*]

1. Radiology and Imaging Sciences, Clinical Center, National Institutes of Health, USA.

2. Artificial Intelligence Resource, National Institutes of Health, USA.

3. Urology Oncology Branch, National cancer institutes, National Institutes of Health, USA.

4. Pathology Department, National Cancer Institutes, National Institutes of Health, USA.

**Corresponding author:**

Ashkan A. Malayeri, MD

**Address:**

Clinical Center, National Institutes of Health

10 Center Drive, 1C352

Bethesda, Maryland 20892

Email: ashkan.malayeri@nih.gov

Phone: (301) 451-4368

**Abbreviations:**

**YOLO:** You Only Look Ones

**PPV:** Positive Predictive Value

**RCC:** Renal Cell Carcinoma

**mAP:** mean Average Precision

**Abstract:**

**Introduction**

This study explores the use of the latest You Only Look Once (YOLO V7) object detection method to enhance kidney detection in medical imaging by training and testing a modified YOLO V7 on medical image formats.

**Methods**

Study includes 878 patients with various subtypes of renal cell carcinoma (RCC) and 206 patients with normal kidneys. A total of 5657 MRI scans for 1084 patients were retrieved. 326 patients with 1034 tumors recruited from a retrospective maintained database, and bounding boxes were drawn around their tumors. A primary model was trained on 80% of annotated cases, with 20% saved for testing (primary test set). The best primary model was then used to identify tumors in the remaining 861 patients and bounding box coordinates were generated on their scans using the model. Ten benchmark training sets were created with generated coordinates on not-segmented patients. The final model used to predict the kidney in the primary test set. We reported the positive predictive value (PPV), sensitivity, and mean average precision (mAP).

**Results**

The primary training set showed an average PPV of $0.94 \pm 0.01$, sensitivity of $0.87 \pm 0.04$, and mAP of $0.91 \pm 0.02$. The best primary model yielded a PPV of 0.97, sensitivity of 0.92, and mAP of 0.95. The final model demonstrated an average PPV of $0.95 \pm 0.03$, sensitivity of $0.98 \pm 0.004$, and mAP of $0.95 \pm 0.01$.

**Conclusion**

Using a semi-supervised approach with a medical image library, we developed a high-performing model for kidney detection. Further external validation is required to assess the model's generalizability.

**Introduction:**

With the advancement in medical imaging technology, the ability to detect anatomical structures has become more accurate, timely, and efficient. Cross-sectional imaging modalities such as computed tomography (CT) and magnetic resonance imaging (MRI) provide valuable information regarding diagnosis, classification and treatment options for kidney cancer (1). In 2021, kidney cancer accounted for 4-5% of all new cancer diagnoses in the United States, and the incidence has been rising for the past three decades (2, 3). The increasing incidental detection of renal masses is most likely due to the increase in the use of cross-sectional imaging (4). In addition, incidentally identified renal masses, often indolent or benign lesions, have led to further use of abdominal imaging for surveillance (5).

Integrating deep learning algorithms such as convolutional neural networks (CNN) has improved the annotation of medical image segmentation. Though there is great benefit in using CNNs, they require significant time and effort to create manually annotated training images. To reduce this initial workload, semi-supervised learning (SSL) methods have enabled the use of a small set of manually annotated images with the ability to apply that to an existing large set of unlabeled data (6).

You Only Look Once—version 7 (YOLOv7) is a deep learning algorithm used for object detection, with the ability to utilize SSL, with speed and accuracy. Several studies have utilized deep learning algorithms like YOLOv3 for object detection of pathologies such as retinal breaks and colonic polyps (7-10). However, in our study, we aim to evaluate the accuracy of YOLOv7 for detecting kidney parenchyma. In this study, we developed and evaluated a model for kidney detection using YOLOv7 semi-supervised learning algorithm.

**Methods:**

*Patient cohort*

This retrospective study examined MRIs from 1084 patients with renal masses who underwent partial or radical nephrectomy or active surveillance between January 2003 to June 2022. two hundred and six patients did not have renal mass. An IRB authorization was obtained for patient recruitment, and signed informed consent was acquired.

A total of 5657 MRI scans from different time points were included. Five different types of scanners were used to capture images. Additional technical information on the MRI scanners is available in supplementary table 1. The following sequences were performed on the patients: multiplanar T2, pre-contrast T1, and post-contrast T1. Post-contrast images were performed in corticomedullary (20-second), nephrogenic (70-second), and excretory (3-minute) phases after contrast material administration.

Out of the total patients in the study, kidneys were segmented in a subset of 223 patients on the excretory phase using ITK-SNAP (version 3.8) by two postdoctoral radiology research fellows. An abdominal radiologist confirmed all segmentations with MRI fellowship training (AAM, 14 years of experience). All segmentations were converted to bounding boxes using preprocessing codes. [https://github.com/Translationalimaginglab/YOLOV7-RCC]

*Image Preprocessing*

All 3D DICOM images were converted to Nifti format, and each slice was separated and saved. Each 3-minute post-contrast image was identified and down-sampled to 1 mm x 1 mm x 1 mm axial slices. Using Rician normalization, all MRIs were normalized. Examples of these MRIs and ground truth segmentations can be seen in *Figure* 1.

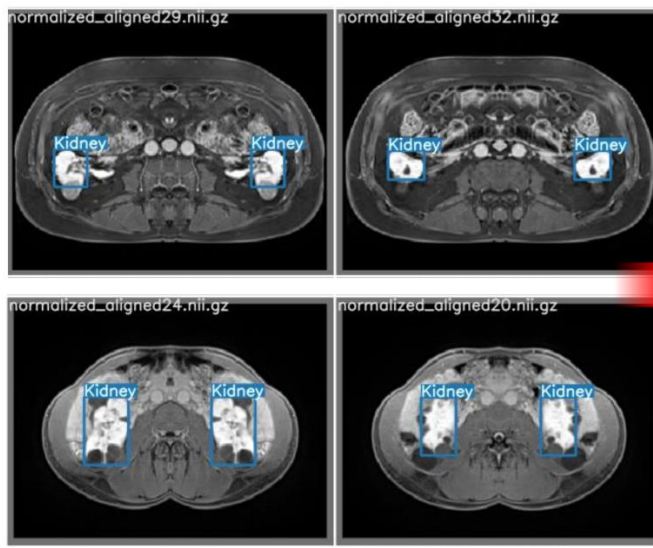
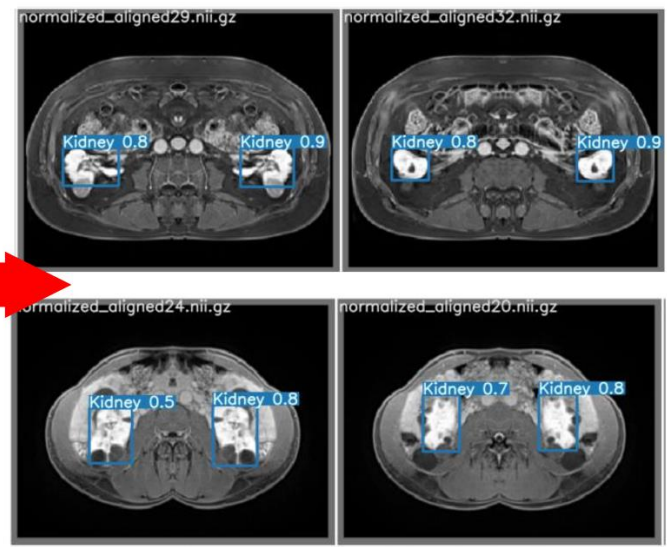

*Figure* **1. Kidney ground truth and related performance images. A.** The ground truth is manually produced. B. Detection performance using the primary model.

*YOLO V7 modification*

By courtesy of Chien-Yao Wang et al. we used the YOLO V7 code and modified it to enable the machine to read, write and predict the objects in '.nii','.nii.gz', and '.dcm' formats. The modified code is stored in our lab's GitHub repository (11).

*Model production*

**Step 1. Primary model:** We trained ten benchmarks of segmented patient scans using YOLOV7. In this training and test set, 80% of patients were randomly assigned to the training group and 20% to the test group. After the model's creation, positive predictive value (PPV), sensitivity, and mean average precision (mAP) for the performance of the models were evaluated, and the best model was chosen.

**Step 2. Detect the object in the dataset:** Detection on all additional non-segmented images was conducted using the chosen model with the best performance in step 1. The corresponding detected kidney coordinates were stored as text files.

**Step 3. Train on all scans:** Another model was trained using the bounding boxes, which we defined in step 2. A total of 861 patients were used in this step to train the model. Step 1 test sets were employed in this set, and any linked scans to those test patients were eliminated from the train set. For training in step 3, weights from the primary model were used to reduce possible false positive training. Model performance was recorded and reported separately as the final results.

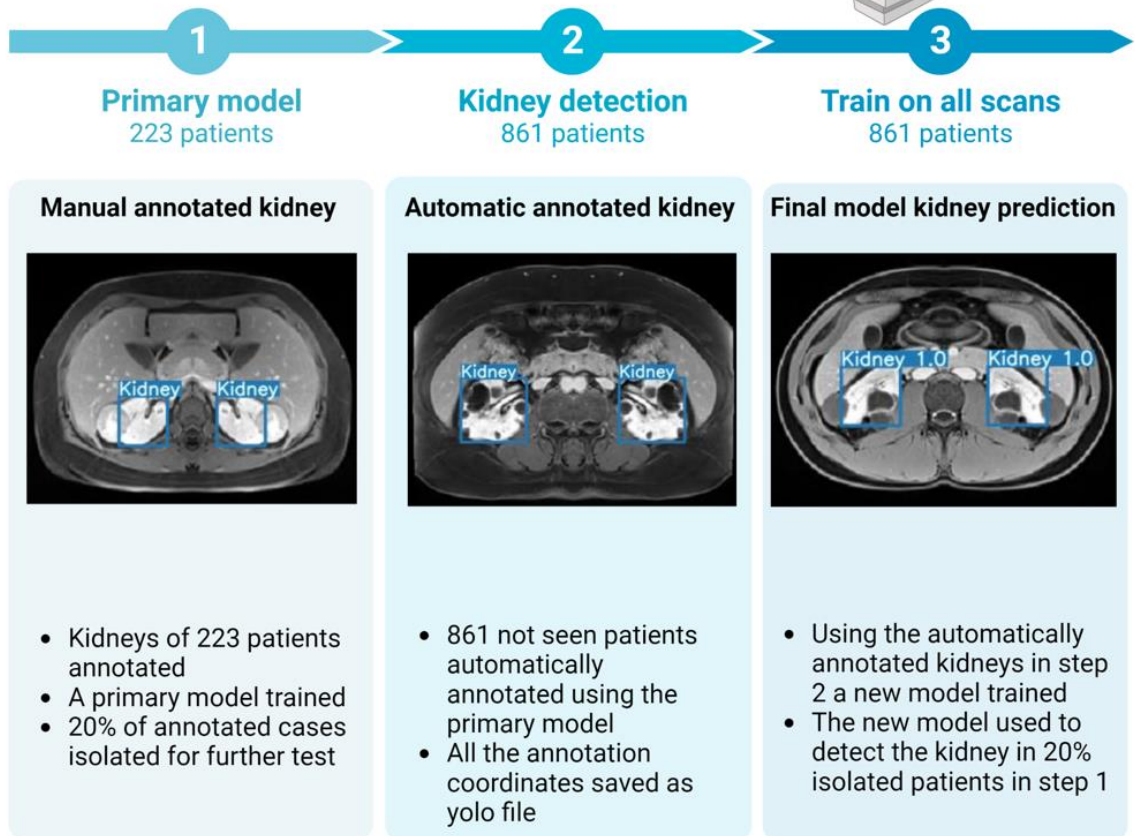

**Figure 2. Study fellow diagram.** In this diagram you can see the process of study step by step.

The learning rate was set at 0.001, and the batch size was 120. The categorical cross-entropy was optimized using the Adam optimizer, with momentum set to 0.9 and weight decay at 0.00005 (12). For each epoch, model weight values were constructed (iterations through the entire data-set). After 100 epochs, the training was terminated due to the lack of further progress in cross entropy and accuracy (list of used hyperparameters added to supplementary).

**Results:**

*Demographic*

Out of the total 1084 patients with renal cell carcinoma included in the study, 57% were male. The median age of the population was 57 (mean = 55.3 ± 14.7). While all patients had at least one renal mass, 79.6% of patients had available pathology for histologic concordance (Table 2).

*Table 1*. **Demographic characteristics of the database.**

| Variable | | Number (%) |
|---|---|---|
| **Gender** | Male | 610 (57) |
| | Female | 456 (43) |
| **Kidney situation** | Normal | 447 (41.2) |
| | **Diseased (RCC*)** | 222 (20.4) |
| | | **Mean ± standard deviation** |
| **Age** | | 55.3 ± 14.7 |

*Model's performance*

The best performance for the primary model had a 0.97 PPV and 0.92 sensitivity with a 0.95 mAP which we used to detect the kidney in the rest of the unsegmented scans. The PPV-sensitivity diagram related to the primary model is demonstrated in **Figure 2.** The best final performance was for the same benchmark, with a PPV of 0.99, sensitivity of 0.99, and mAP of 0.98 (**Table 3**).

*Table 2*. **Best performance for each trained and tested set.** In this table, each benchmark performance is demonstrated. The best performance model for detecting tumors in the entire image database was the second benchmark with 0.97 PPV and 0.91 sensitivity.

| Benchmark number | Best primary model test performance | | | Best all scans test performance | | |
|---|---|---|---|---|---|---|
| Benchmark number | PPV | Sensitivity | mAP at 0.5 confidence | PPV | Sensitivity | mAP at 0.5 confidence |
| 1 | 0.93 | 0.87 | 0.91 | 0.98 | 0.98 | 0.97 |
| 2 | 0.97 | 0.91 | 0.95 | 0.99 | 0.98 | 0.96 |
| 3 | **0.97** | **0.97** | **0.95** | **0.99** | **0.99** | **0.98** |
| 4 | 0.93 | 0.83 | 0.88 | 0.96 | 0.98 | 0.95 |
| 5 | 0.93 | 0.85 | 0.90 | 0.97 | 0.98 | 0.97 |
| 6 | 0.93 | 0.86 | 0.88 | 0.98 | 0.98 | 0.95 |
| 7 | 0.92 | 0.85 | 0.87 | 0.95 | 0.97 | 0.92 |
| 8 | 0.93 | 0.85 | 0.90 | 0.91 | 0.98 | 0.95 |
| 9 | 0.93 | 0.87 | 0.91 | 0.9 | 0.98 | 0.94 |
| 10 | 0.95 | 0.88 | 0.92 | 0.92 | 0.98 | 0.96 |
| Average performance | 0.94 ± 0.01 | 0.87 ± 0.04 | 0.91 ± 0.02 | 0.95 ± 0.03 | 0.98 ± 0.004 | 0.95 ± 0.01 |

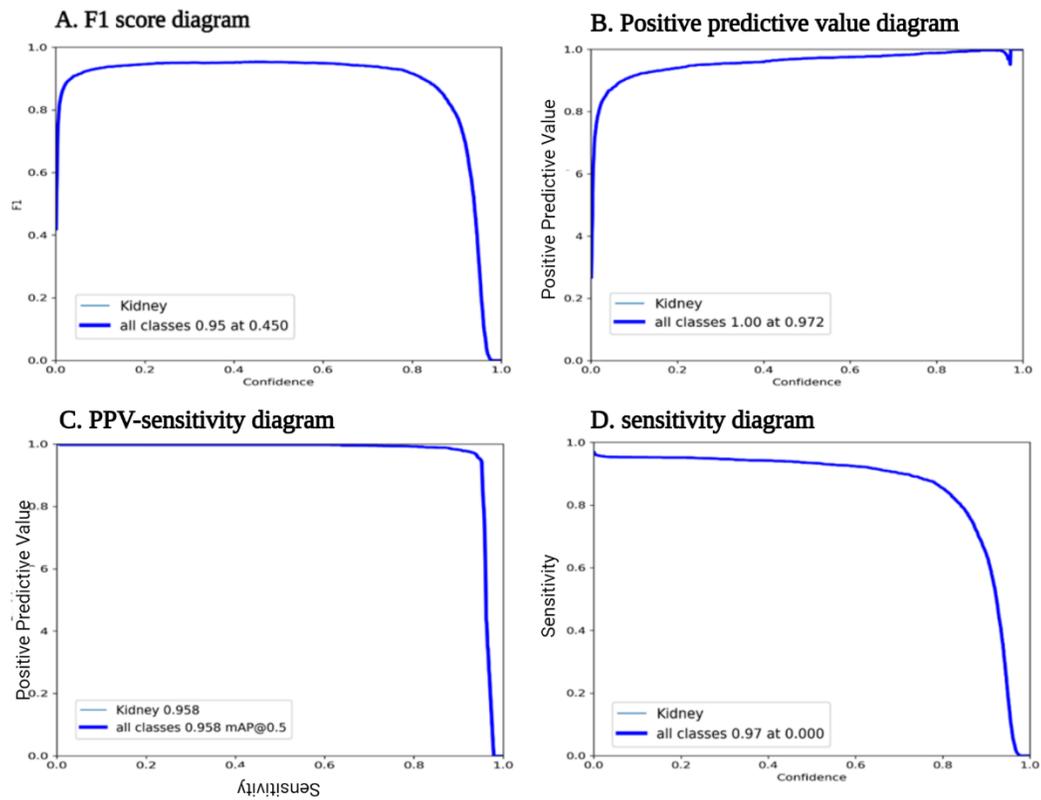

*Figure 3*. **Primary best model performance diagrams. A.** F1 score of 0.95 at 0.45 confidence was calculated. **B.** PPV was 0.97 **C.** PPV-Sensitivity curve showed that mAP of 0.5 was 0.95 **D.** In this diagram sensitivity showed to be 0.97.

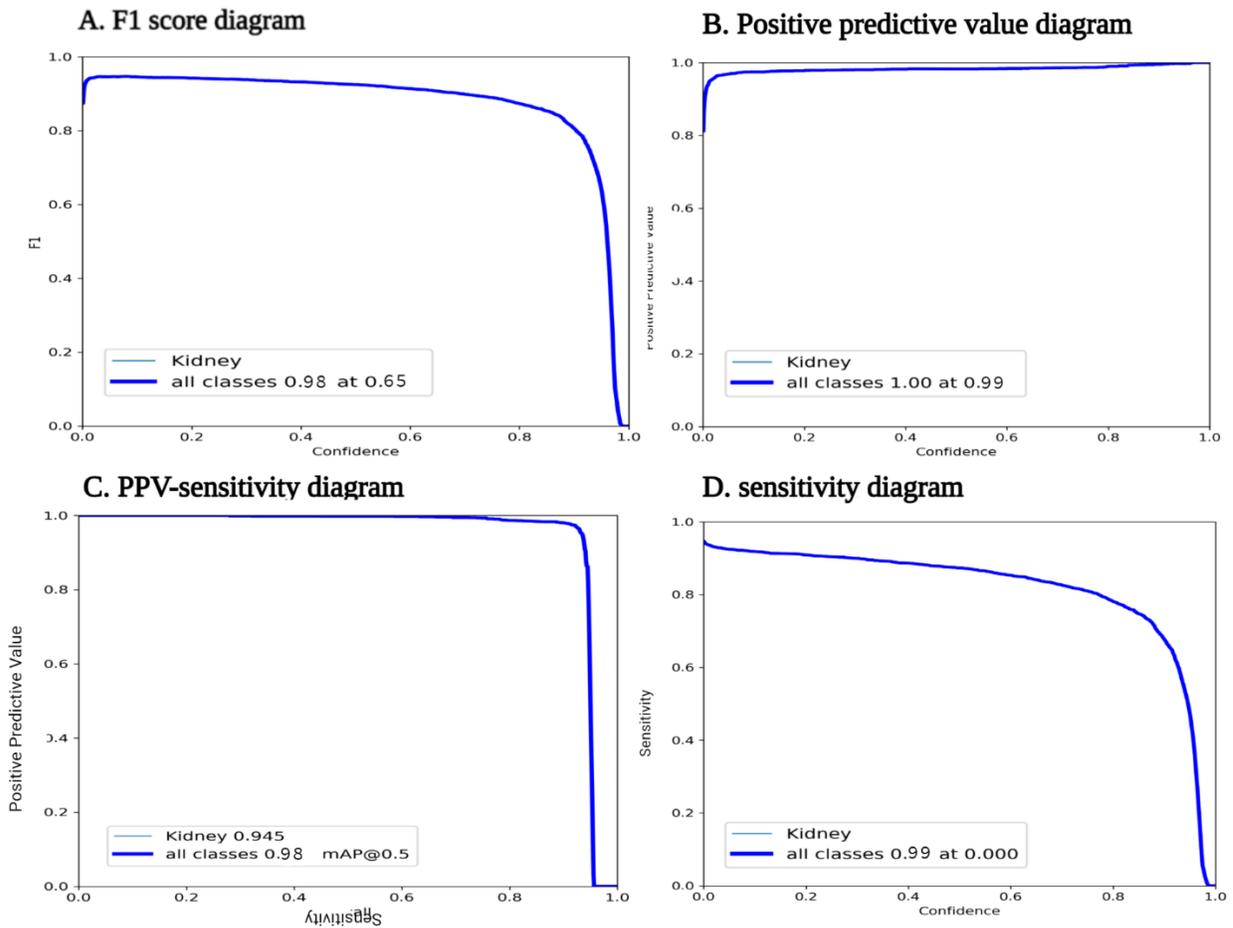

*Figure 4.* **Final best model performance diagrams. A.** F1 score of 0.98 at 0.65 confidence was calculated. **B.** PPV was 0.99 **C.** PPV-Sensitivity curve showed that mAP of 0.5 was 0.98 **D.** In this diagram sensitivity showed to be 0.99.

**Discussion:**

Currently, detection methods are routinely used to identify anatomical and anomalous features and prostheses in the human body. In addition to their use in detection, segmentation methods are expanding quickly to improve the assessment of volumetric structure and biological activity. This study showed that detecting renal parenchyma using YOLO V7 with the medical imaging formats is possible and can succeed. The model we trained in this study can facilitate early diagnosis and monitoring by enabling early detection of kidney anomalies, which is crucial for timely intervention and treatment. It can also be employed in monitoring the progression or regression of renal conditions, allowing physicians to adapt treatment plans as necessary. Additionally, the model can be used for quantitative assessment of renal function by analyzing renal parenchymal volume, serving as an indicator of kidney function and helping in diagnosing and managing conditions like chronic kidney disease, acute kidney injury, and renal transplant patients. By providing precise information about the kidney's structure, our model can assist in personalized treatment planning, helping clinicians design tailored treatment strategies for patients with renal conditions, improving therapeutic outcomes and reducing complications. Furthermore, the model can be utilized in research and clinical trials, enabling researchers to investigate the effectiveness of new drugs or therapies on kidney tissue and facilitating the identification of suitable candidates for clinical trials, potentially accelerating the development of novel treatments.

The Kidney and Kidney Tumor Segmentation Challenge (KiTS19) marked a significant milestone in kidney image processing as the first open-source database. This challenge used computed tomography (CT) scans and advanced the identification and segmentation of kidney and associated malignancies. The highest dice score for tumor and kidney segmentation for KiTS19 was 0.9184. Though we did not segment the lesions and kidney, detecting the kidney in MRI can be challenging, which speaks to the high performance of our approach. Our method, successfully detected renal parenchyma with a final detection performance of 0.98 for mAP, demonstrating its accuracy in finding the kidney parenchyma. This achievement highlights the potential of our model in handling the complexities of MRI-based kidney detection, while also showcasing its adaptability to various medical imaging formats. Furthermore, our model's robust performance contributes to the growing body of evidence supporting the use of advanced machine learning techniques in medical imaging and renal health.

Other study in this field that provided a model for kidney parenchymal segmentation were conducted on normal kidney structures, the study by Taro Langner et al. who examined the feasibility of automatically segmenting the renal parenchyma using the UK Biobank MRIs, involving approximately 40,000 healthy volunteers, with a DICE similarity scale of 0.96 (13).In comparison, our model achieved a final detection performance of 0.98 for mAP, indicating a high accuracy in detecting kidney parenchyma. The difference in performance could be attributed to various factors such as our study employing YOLO V7 for detection, a diverse dataset from different scanners, and focusing on detecting kidney parenchyma rather than segmentation.

However, direct comparison of the two studies might not be entirely appropriate due to differences in task, methodology, and dataset, but both demonstrate promising results in the field of kidney segmentation and detection, paving the way for improved diagnosis and treatment of renal conditions.

An important limitation we can identify in our study is not using an external validation of the model. Using many images (more than 3 million images) and several types of scanners may make it easier to anticipate the outcome of external validation. However, we may still need external validation to confidently verify the model's performance. Another limitation is our model's focus on detecting renal parenchyma and not addressing the detection of other important kidney structures, such as the renal pelvis, calyces, and vasculature. This limitation may impact the model's applicability in assessing various kidney conditions that require the evaluation of these structures. Furthermore, our study did not evaluate the integration of the model into clinical workflows or its impact on clinical decision-making. Further research is needed to understand how the model can be best implemented and used by healthcare professionals to improve patient care.

**Conclusion:**

This study's encouraging findings suggest that the developed model might be used for kidney parenchyma detection. In addition, it demonstrated that modified YOLOv7 code may be used on medical imaging format directly to construct YOLOv7 models. This investigation must also be conducted on scans from external institutions to prove its validity.

**Acknowledgments:**

The authors of this study made use of the advanced computational resources available at the NIH HPC Biowulf cluster (http://hpc.nih.gov). Financial support for this research was provided by the NIH Intramural Research Program, as well as the NIH Clinical Center's Research Award for Staff Clinicians (RASCL). Additional gratitude is extended to the Office of Biomedical Translational Research Informatics (BTRIS), especially Gloria Oshegbo, for their valuable assistance and data provisioning.

Supplementary table 1. Scanners technical informations. In this study scans were captured using 5 different scanners with 1.5 or 3 Tesla magnetic field.

| Company/Machine brand | Achieva | Aera | Avanto | Biograph_mMR | Echelon Oval | Skyra | Skyra_fit | TrioTim | Verio | Grand Total |
|---|---|---|---|---|---|---|---|---|---|---|
| **Hitachi** | | | | | 1 | | | | | 1 |
| 1.5 T | | | | | 1 | | | | | 1 |
| **Philips Medical Systems** | 1621 | | | | | | | | | 1621 |
| 1.5 T | 1044 | | | | | | | | | 1044 |
| 3 T | 577 | | | | | | | | | 577 |
| **SIEMENS** | | 3562 | 1 | 130 | | 2 | 1 | 1 | 338 | 4035 |
| 1.5 T | | 3562 | 1 | | | | | | | 3563 |
| 3 T | | | | 130 | | 2 | 1 | 1 | 338 | 472 |
| **Grand Total** | 1621 | 3562 | 1 | 130 | 1 | 2 | 1 | 1 | 338 | 5657 |